\documentclass[aps, twocolumn]{revtex4-1}
\usepackage{hyperref,amsmath,mhchem,amssymb}
\usepackage{graphicx,float}

\usepackage{balance}
\pdfoutput=1

\begin{document} 
 
\title{Exotic Vortex Lattices in Binary Repulsive Superfluids}
\date{\today}
\author{Luca Mingarelli  and Ryan Barnett}
\affiliation{Department of Mathematics, Imperial College London,
London SW7 2AZ, United Kingdom}
\begin{abstract}
We investigate a mixture of two repulsively interacting superfluids with different constituent particle masses: $m_1 \ne m_2$. Solutions to the Gross-Pitaevskii equation for homogeneous infinite vortex lattices predict the existence of  rich vortex lattice configurations, a number of which correspond to Platonic and Archimedean planar tilings.
Some notable  geometries include the snub-square, honeycomb, kagome, and herringbone lattice configurations.  We present a full phase diagram for the case $m_2/m_1=2$ and list a number of geometries that are found for higher integer mass 
{ratios}. 

\end{abstract}

\maketitle

{
While in ancient times the nature of planar tessellations with regular polygons \cite{tilingsbyregularpolygons} was  contemplated within the context of the arts, today such regular structures are also known to arise from the self-assembly  of systems as diverse as metal alloys \cite{kasper1,kasper2}, DNA nanoparticles \cite{zhang2016self,preisler2017irregular}, liquid crystals \cite{liquidcrystals},
nanoconfined water \cite{nanowater2016}, and a range of systems made of different molecules and polymers \cite{matsushita2008precise,ecija2013five}. 
These regular structures  are important for their resulting mechanical, optical, electrical, and chemical properties.
Appreciating how certain systems self-assemble into specific regular tessellations is relevant both for the 
fundamental understanding of the systems themselves as well as for the design of new metamaterials \cite{minimaldesign,Millan2014,Jovanovic:08}.

The simplest periodic tessellations of the 2D plane are the Platonic tilings, which are formed with only one kind of regular polygon. 
At the next level of complexity are the Archimedean tilings of the plane  \cite{1619HM}.   For this case, two or more regular
polygons are required with the condition that all vertices are equivalent.  
Typically, one finds that simpler and more fundamental systems give rise to simpler structures (e.g.\ Platonic), while Archimedean tilings are usually associated with systems having more complex interactions.
For this reason it is of particular interest to  encounter systems whose constituents are relatively simple but that can still attain structures as complex as the Archimedean tilings, or more. 

In this work, we investigate periodic tessellations of the plane within the context of vortices appearing 
in superfluid systems (identifying the vortex lattice points as vertices of polygons).  
In addition
to the Platonic tilings (triangular, square, and honeycomb), we also find Archimedean tilings
(snub-square and kagome), as well as more complex structures (e.g.\ herringbone).  These periodic lattices are
summarised in Fig.\  \ref{tableofstates}.

A defining characteristic of superfluids follows from their response to mechanical rotation.
Rather than rotate like a rigid body, superfluids will generically nucleate vortices, each carrying
a quantised circulation \cite{Onsager49}.  Moreover, when rotated sufficiently rapidly, a vortex lattice will form  \cite{Feynman55_2}. 
While the first superfluid vortex lattices were observed in liquid helium \cite{vinen58,Deaverquantvortexp}, more recently,
ultracold atoms have provided an arena to explore the physics of superfluid vortices with great precision and control 
\cite{Fetter09}.

{
It has long been appreciated that the triangular vortex-lattice configuration is  
robust and stable for experimentally relevant parameters in a} single-species superfluid \cite{Kleiner64, tkachenko1966vortex}. In search for {richer} vortex crystal structures, 
{ multicomponent superfluids under rotation have been considered, including
spinor condensates \cite{Kawaguchi12} and binary mixtures of condensed atomic gases
\cite{Mueller02, Kasamatsu03, Barnett10, Mason11, Kuopanportti12,Keifmmode06,uranga2018infinite}. 
 A common approach to solve the associated equation of motion, the Gross-Pitaevskii equation, consists in projecting the wave function $\psi$ onto the lowest Landau level (LLL) \cite{Abrikosov57,Kleiner64} (a method of including corrections from higher Landau levels
is presented in Ref.~\cite{Kita}). { This is appropriate in the limit where the magnetic length
 $\ell_B=\sqrt{\frac{\hbar}{2m\Omega}}$ (defined in analogy with quantum Hall systems) is { small in comparison
to  the} healing length $\xi=\sqrt{\frac{\hbar^2}{2mg\bar{\rho}}}$ (here $\bar{\rho}$ is the average superfluid density). 
 Although capable of producing very precise solutions, and accurately classifying different phases in binary superfluid systems \cite{Mueller02}, it is experimentally restrictive to consider the LLL only. }

{The other commonly employed approach \cite{Kasamatsu03,Barnett10,Mason11,Kuopanportti12}  consists of
computationally minimising the energy of a system of many vortices confined by a trapping potential.}
 For such systems, the confining potential works to obscure the underlying vortex lattice. Therefore, in order to infer the ideal periodic configuration of vortices, one must require $\ell_B$ to be much smaller than the condensate size.  In practice, this means one must computationally address systems with at least around 100 vortices. Such large systems become increasingly complicated to solve, both because the energy landscape acquires more degrees of freedom and because more computational points are required to accurately describe the vortex lattice.  

Recently, a method has been proposed which overcomes the drawbacks arising from both of the above  approaches \cite{Noi}, which consists of seeking the lowest energy solutions to the Gross-Pitaevskii equation in a quasiperiodic unit cell by means of the so-called {\it Magnetic Fourier Transform} which gives a straightforward diagonalisation of the relevant linear operators of the model. The generalisation of this method to a multicomponent system was given in \cite{Newarticle}, where the full characterisation of the phase diagram for a binary system with constituents of equal masses was obtained. 
The method addresses infinite systems, enabling one to investigate universal aspects of vortex lattices without the (typically small) reconstruction resulting from confining potentials that varies from experiment to experiment.


\label{Sec:2}
Within Gross-Pitaevskii mean field theory, the energy functional associated with a two-species system with masses $m_j$, in a rotating frame of reference, can be written as ${E=\int\mathcal{E}[\psi_1,\psi_2]\text{d} x\text{d} y}$ with energy density
  \begin{equation}\begin{split}\label{GPEnergy}
\mathcal{E}[\psi_1,\psi_2]=&\sum_{j=1}^2 \Bigl[\frac{\hbar}{2m_j}|\nabla\psi_j|^2+\frac{1}{2} m_j\omega_j^2r^2|\psi_j|^2\\
&-\psi_j^\dagger\Omega L_z\psi_j-\mu_j|\psi_j|^2\Bigr]+\frac{1}{2}{\boldsymbol{\rho}^T \mathcal{G}\boldsymbol{\rho}}.
\end{split}
\end{equation}
The matrix $\mathcal{G}$ accounting for intra- and interspecies interactions, required to be positive semidefinite in order to ensure miscibility of the two superfluids,
is defined as
\begin{equation}
\mathcal{G}=\begin{pmatrix}
    g_{1} & g_{12} \\
    g_{12}  & g_2 \\
    \end{pmatrix},
\end{equation}
and its elements can be related to the $s$-wave scattering lengths $a_{jk}$: $g_j=4\pi\hbar^2a_{jj}/m_j$,
$g_{12}=2\pi\hbar^2a_{12}(m_1+m_2)/m_1m_2$. Moreover, $\boldsymbol{\rho}^T=(|\psi_1|^2,|\psi_2|^2)$ { and} $L_z=-i\hbar(x\partial_y-y\partial_x)$ {where} $\Omega$ is the rotational frequency, and $\mu_j$, $\omega_j$ are, respectively, the chemical potential and the trapping frequency of the $j$th species. 
Finally, let us introduce here the dimensionless parameter 
\begin{align}
\alpha=g_{12}/\sqrt{g_1g_2},
\end{align}
 quantifying the interspecies interaction strength relative to the intraspecies strengths: this will prove useful in the following. The positive semidefiniteness of $\mathcal{G}$ can now be expressed more compactly as $\alpha\le 1$.

To find the ground state of Eq.~(\ref{GPEnergy}), we employ the method described in  Ref.~\cite{Noi}
 and extended to multicomponent systems in Ref.~\cite{Newarticle}.    The method 
 consists of approaching the discretisation of Eq.~\eqref{GPEnergy} through the introduction of a generalised nonlinear Hofstadter model, which reduces to the  model \eqref{GPEnergy} in the continuum limit. The main advantage of this approach, is that it preserves the  gauge symmetries of the system, introduced into the Hofstadter model by Peierls substitutions \cite{Peierls33}. This further permits  a direct diagonalisation of relevant terms in the model by means of a  {\it Magnetic Fourier Transforms} \cite{Noi}. Finally one can obtain a piecewise diagonalised model, which is invariant on the choice of the gauge; the resulting linear and nonlinear terms can thus be split, propagation achieved at the desired level of accuracy in time through appropriate split-step methods, and the lowest-energy solutions obtained by propagation in imaginary time $\tau=it$.

\label{Sec:3}

{The states attainable in a system with equal constituent masses $m_1=m_2$ were investigated originally in Ref.~\cite{Mueller02}.
The results were extended beyond the lowest-Landau level regime  in Ref.~\cite{Newarticle}, and the resulting possible vortex lattices for $m_1=m_2$   are summarised in Fig.~\ref{tableofstates}. These consist of triangular, oblique, square, and rectangular vortex lattices. 

The lift of the requirement of equal masses enables a number of additional geometries to arise, most importantly because the average
vortex densities of the two superfluids are no longer equal. Specifically, from a relation due to Feynman \cite{Feynman55_2}, the ratio of vortex densities is
\begin{align}
\label{feynman}
\frac{\rho_v^{(1)}}{\rho_v^{(2)}} = \frac{m_1}{m_2}
\end{align}
where $\rho_v^{(j)}$ is the density of vortices for the $j$th species.
When there is no coupling between the superfluids, two decoupled triangular lattices will form.  On the other 
hand, when $g_{12} \ne 0$, vortices of the different species will interact.}
Because of this, we are presented with a new range of exotic and complex vortex lattices. Before going into a discussion of main results, a brief comment on the commensurability of the system under study is in order.


When considering components whose constituents have equal masses, one can ignore an issue that presents itself when considering infinite vortex lattices in the more general case, namely, that of commensurability. When $m_1=m_2$, one obtains a variety of commensurate vortex lattices as a function of the interspecies interaction strength, $g_{12}$.
 Since we are interested in studying systems with $m_1 \ne m_2$, we will briefly consider some issues related to commensurability at a rather general level first. Without loss of generality, in the following we will assume $m_2>m_1$.

When the interspecies interaction is zero, the rotating system will form two independent triangular lattices oriented in an arbitrary way with respect to one another.  
Denoting the vortex lattice vectors of the $i$th component as ${\bf b}_1^{(i)}$ and ${\bf b}_2^{(i)}$, the two lattices are given by
the collection of points 
$\Lambda_i = \{ n {\bf b}_1^{(i)} +  \ell {\bf b}_2^{(i)} | n,\ell \in  \mathbb{Z} \}$.  In this, without loss of generality, we arranged the lattices so that they both have a point in common at the origin.  For the lattices to be commensurate, $\Lambda_1$ and $\Lambda_2$ must share an infinite number 
 of points.  Recalling Eq.~(\ref{feynman}), this
means that there must be solutions to the following Diophantine equation:
$m_2 (n^2 + \ell^2 + n \ell) = m_1 (p^2 + q^2+ pq)$, where $n,\ell,p,q$ are integers.  In the following we restrict to cases where $m_2/m_1$
is an integer, as we only consider these cases later in this work.  It can be seen that solutions to the Diophantine equation other than the trivial one, exist
if and only if the mass ratio is a 
L\"oschian number \cite{loschian}, namely, if it can be expressed as
\begin{equation}
\frac{m_2}{m_1}=\mu^2+\mu\nu+\nu^2=1,3,4,7,9,12,13,...
\end{equation}
with $\mu,\nu\in\Bbb{Z}$. 
When this condition is satisfied, one vortex lattice will be found rotated with respect to the other by an angle ${\theta=\arctan{\left(\frac{\sqrt{3}\nu}{2\mu+1}\right)}}$.
Thus we find $\theta=0$ when the mass ratio is a perfect square. 
More generally, when the mass ratio is not restricted to integer values, 
the general condition for commensurability is for $m_2/m_1$ to be a ratio of any two L\"oschian numbers.

For small $g_{12}$, where we expect two triangular lattices, this result implies that we will not be able to find commensurate ground states for mass ratios which are not L\"oschian. However we might still be able to find commensurate states when there are stronger interactions between components. This is indeed what we find for the simplest non-L\"oschian case of $m_2/m_1=2$.

From a computational point of view, it is not straightforward to identify { incommensurate} states. This is because the simulated states are periodic in the density by construction (having the periodicity of the chosen computational unit cell).
In practice, we check whether the same state configuration is obtained for larger unit cells.  If the vortex lattice changes upon continually increasing the size of the unit cell by integer multiples, the corresponding state is concluded to be incommensurate.

\begin{figure}[!]
\includegraphics[width=\linewidth]{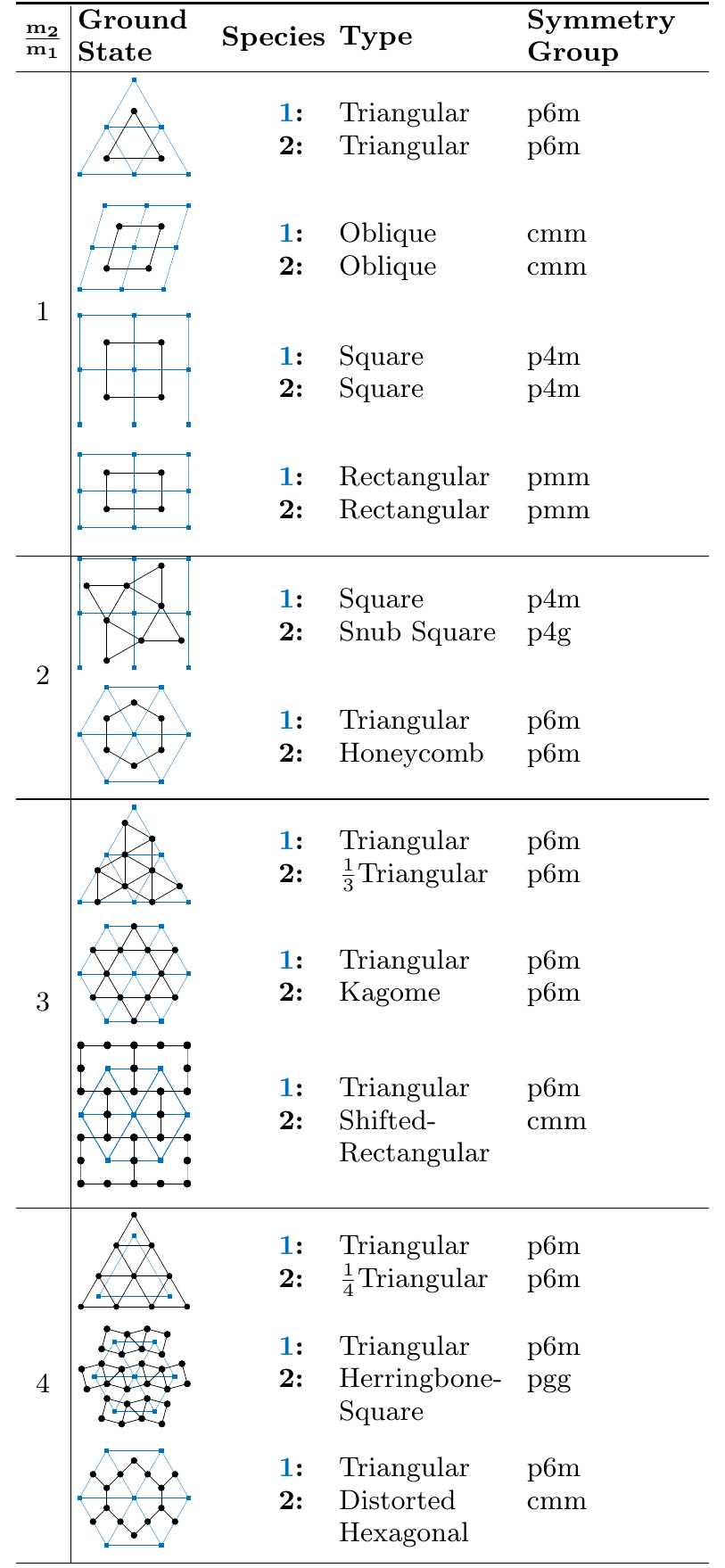}
\caption{Regular commensurate ground state configurations for different mass ratios.  
In the final column, the lattice symmetry classification is given in the IUC convention.
We denote by ``$\frac{1}{n}$Triangular'' the triangular lattice with unit cell one $n$th of the unit cell area of the  lattice denoted as ``Triangular''.}
\label{tableofstates}
\end{figure}


Despite $m_2/m_1=2$ not being L\"oschian, and thus having two incommensurate triangular lattices for ${\alpha\approx 0}$, we find other states which are commensurate for higher interspecies interactions. This mass ratio is of particular relevance since
 in experiments one can achieve it to a  good approximation with the mixture of isotopes
\mbox{\ce{^{41}_{}K}--\ce{^{87}_{}Rb}} (with mass ratio ${m_2/m_1\approx 2.1}$)\cite{Modugno02,Ferrari02}, but also in principle with
\mbox{\ce{^{87}_{}Rb}--\ce{^{174}_{}Yb}} ($m_2/m_1 \approx 2.0014$) and 
\mbox{\ce{^{84}_{}Sr}--\ce{^{168}_{}Er}} ($m_2/m_1 \approx 2.0013$).

Requiring $\ell_B/\xi$ to be the same for both components, the ground state phase diagram associated with the mass ratio $m_2/m_1=2$ is presented in Fig.\ \ref{PD}.
In this scenario we encounter two new commensurate ground states. For $\alpha=0$, when the two species are not interacting, 
 two incommensurate triangular lattices are formed. The first transition we find when increasing $\alpha$  transforms the ground state
 of the lighter species into a square lattice, while the heavier is transformed into a snub-square lattice. 
The second transition is second order: the square vortex lattice associated with the lighter species transforms into a triangular lattice while the snub-square lattice in the heavier component is transformed into a honeycomb lattice. 

\begin{figure}[!t]
\includegraphics{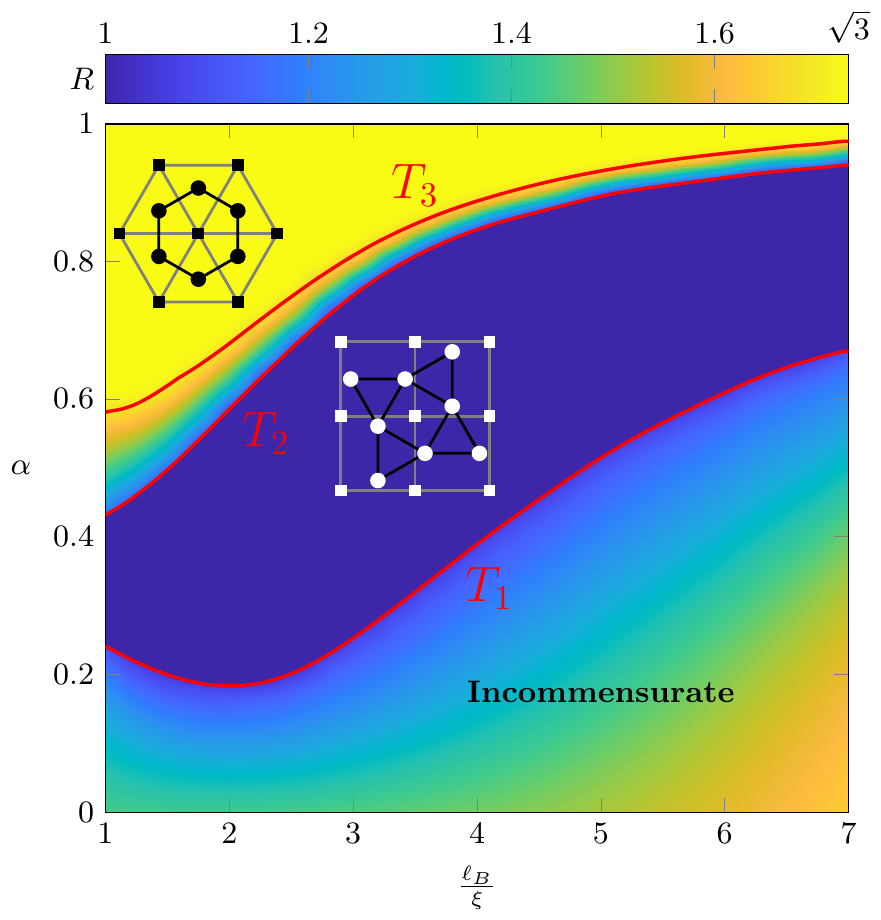}
\caption{ 
Phase diagram describing the ground states of two interacting superfluids with equal particle number per unit cell $\mathcal{N}_1=\mathcal{N}_2$ and mass ratio $m_2/m_1=2$.  In the area above $T_3$ a honeycomb lattice of  vortices is formed in the heavier species, interlaced with the triangular lattice which forms in the lighter species. Between $T_3$ and $T_2$ the system undergoes 
a continuous transformation to  a snub-square lattice for the heavier component interlaced with a square lattice of vortices in the lighter component. This configuration remains stable in the region enclosed by $T_2$ and $T_1$. Finally, as expected from the earlier arguments, an incommensurate state is found for small values of  $\alpha$. 
	}
	\label{PD}
\end{figure} 



Figure~\ref{tableofstates} provides a summary of the commensurate ground states we find at higher mass ratios. The next integer case, namely, that of mass ratio $m_2/m_1=3$, might be  realisable with isotopes
\ce{^{41}_{}K}--\ce{^{133}_{}Cs} ($m_2/m_1 \approx 3.2$)\cite{mixKCs}, for the mixture
\ce{^{7}_{}Li}--\ce{^{23}_{}Na} ($m_2/m_1 \approx 3.3$), or for
\ce{^{52}_{}Cr}--\ce{^{164}_{}Dy} ($m_2/m_1 \approx 3.1561$).
In this case, as for the case $m_2/m_1=1$ \cite{Newarticle} one finds  a complete commensurate phase diagram. For $\alpha\approx 0$ we find two commensurate triangular lattices, one tilted with respect to the other by an angle $\theta=\pi/6$. With a stronger interaction we find at first that the lighter component's vortices form a triangular lattice, while the heavier component arranges its vortices into a kagome lattice.
For an even higher interspecies interaction, closer to the misciblity-immiscibility boundary $\alpha=1$, the latter turns into a shifted-rectangular lattice, while the lighter component retains its triangular arrangement.

Finally, the last mass ratio we consider is ${m_2/m_1=4}$, which could be implemented with the isotopes 
\mbox{\ce{^{23}_{}Na}--\ce{^{87}_{}Rb}} that have mass ratio $m_2/m_1 \approx 3.8$ \cite{mixNaRb} or more accurately with 
\mbox{\ce{^{41}_{}K}--\ce{^{164}_{}Dy}} ($m_2/m_1 \approx 4.0021$).
As mentioned, because this mass ratio is a perfect square, at $\alpha\approx 0$ we can obtain two commensurate triangular lattices tilted with respect to each other by an angle $\theta=0$. 
When  $\alpha$ is increased, the symmetry is broken along one direction and we observe the formation of a new family of states made of rectangles centered on the vortices belonging to the triangular lattice formed by the lighter species, and arranged 
in a herringbone configuration.
Finally, a further increase of the interaction parameter $\alpha$ leads to a lattice made of nonregular hexagons centered on the triangular lattice of the lighter species.

In the limit of large mass ratios, we expect the vortex lattices associated with the lighter component to be triangular at all miscibile interspecies interaction regimes; the vortices in the heavier component will then be found arranged around the lighter triangular lattice so as to minimise the total energy.   In this regime, the effective interactions experienced by the heavier component's vortices is suppressed by a factor of $\sim m_1/m_2$ in comparison with the intraspecies interactions of the vortices in the lighter component.  
We can indeed observe this pattern in Fig.~\ref{tableofstates}, and expect regular structures built on top of the triangular lattice for higher mass ratios. One such example, not included in Fig.~\ref{tableofstates}, is that of the { snub-trihexagonal} lattice, which we find in binary systems with mass ratio $m_2/m_1=6$. 
Notice in particular, that while most of the configurations considered so far are achiral, the snub-trihexagonal lattice is chiral. The oblique lattice in particular, as well as other transitory states such as the intermediate states in between snub-square and honeycomb, or between the kagome and the shifted-rectangular configurations are chiral as well.
It is worth pointing out that all the results discussed so far have been obtained under the assumption of zero temperature. At small but finite temperatures we expect small deviations from our results, although these might still be experimentally relevant. In particular, one can expect thermal fluctuations to suppress phase separation \cite{Roy_Angom}  with a likely consequence of an upwards shift of the transition boundaries in Fig~\ref{PD}. Such deviations could be more accurately taken into account by employing an extension of our method to account for the Zaremba-Nikuni-Griffin theory \cite{griffin2009bose, Lee_Proukakis}. 
Finally, let us make one more comment on the nature of the geometrical configurations obtained. One question which might naturally arise is whether these vortex lattices are specific to the Gross-Pitaevskii theory, or whether they could be similarly obtained in systems of particles with isotropic interactions. It is true that complex configurations such as the snub-square and kagome lattices can be obtained in isotropic systems a well. This is, however, usually achieved with finely tuned potentials obtained by reverse engineering methods 
\cite{Edlund_Nilsson}.
%

 
In conclusion, we have employed the method presented in Refs.~\cite{Noi,Newarticle} to investigate binary homogeneous systems of superfluids 
where the two species of constituent particles have unequal masses.
Exotic vortex configurations naturally arising from the minimisation of the energy  were found. Interestingly, some configurations correspond to Archimedean tilings of the plane, such as the snub-square and kagome lattice configurations occurring for mass ratios  $m_2/m_1=2$ and $3$, respectively.  Additional periodic lattices corresponding to nonregular polygons were also found.
We hope these results will foster further work and investigation: more exotic configurations are to be expected at different mass ratios
(including those taking on fractional values), and a comprehensive classification of the achievable geometries is likely attainable through further research.    
Moreover it is worth noticing that here we have enforced the conservation of individual total particle numbers $\mathcal{N}_1$ and $\mathcal{N}_2$. This leads to a global $U(1)\times U(1)$ symmetry which is broken when enforcing the conservation of $\mathcal{N}=\mathcal{N}_1+\mathcal{N}_2$ instead \cite{RADZIHOVSKY20082376}. The context of spinor condensates in which this symmetry breaking is relevant, would likely lead to different phase diagram configurations. 
While negative interspecies interactions are more common in condensate mixtures, it is also possible to have attractive interactions.  Infinite vortex lattices for $\alpha<0$ warrant further consideration.
Another interesting open question is if quasicrystalline vortex lattices are possible in such superfluid mixtures.
We hope to see the vortex lattices reported in this work realised, as they are within current experimental capabilities.


\begin{acknowledgments}
We are grateful to Eric Keaveny for past collaborations and useful discussions.
This work was supported in part by the European Union's Seventh Framework Programme for
research, technological development, and demonstration
under Grant No.\ PCIG-GA-2013-631002.
Some of this work was done  at the Aspen Center for Physics, which is supported by National Science Foundation grant PHY-1607611 and the Simons Foundation.
\end{acknowledgments}

\bibliographystyle{apsrev}

\end{document}